\documentclass[12pt]{article}
\usepackage{amsfonts,amssymb,mathrsfs}
\usepackage{hyperref}
\newcommand{\beq}{\begin{equation}}
\newcommand{\eeq}{\end{equation}}

\newcommand{\pt}{\partial}

\newcommand{\bt}{\beta}
\newcommand{\g}{\gamma}
\newcommand{\de}{\delta}
\newcommand{\varep}{\varepsilon}
\newcommand{\s}{\sigma}
\newcommand{\la}{\lambda}
\newcommand{\mc}{\mathcal}

\newcommand{\un}{\underline}

\begin{document}
\begin{center}
{\Large\textbf{Massless spinning particle and null-string on $AdS_d$: projective-space approach}}\\[0.3cm]
{\large D.V.~Uvarov\footnote{E-mail: d\_uvarov@\,hotmail.com}}\\[0.2cm]
\textit{NSC Kharkov Institute of Physics and Technology,}\\ \textit{61108 Kharkov, Ukraine}\\[0.5cm]
\end{center}
\begin{abstract}
Massless spinning particle and tensionless string models on $AdS_d$ background in the projective-space realization are proposed as constrained Hamiltonian systems. Various forms of particle and string Lagrangians are derived and classical mechanics is studied including the Lax-type representation of the equations of motion. After that transition to the quantum theory is discussed. Analysis of potential anomalies in the tensionless string model necessitates introduction of ghosts and BRST charge. It is shown that quantum BRST charge is nilpotent for any $d$ if coordinate-momentum ordering for the phase-space bosonic variables, Weyl ordering for the fermions and $cb$ ($\g\bt$) ordering for ghosts is chosen, while conformal reparametrizations and space-time dilatations turn out to be anomalous for the ordering in terms of positive and negative Fourier modes of the phase-space variables and ghosts.
\end{abstract}

\setcounter{equation}{0}
\def\theequation{\thesection.\arabic{equation}}
\section{Introduction}

The study of field dynamics in (anti-)de Sitter and Minkowski spaces viewed as the manifolds embedded into flat space-times with extra dimension(s) was initiated in the seminal works of Dirac \cite{Dirac}. In the context of point-particle mechanics the embedding of $d-$dimensional anti-de Sitter space into $(d+1)-$dimensional flat space-time with metric signature $(2,d-1)$ parametrized by the homogeneous coordinates on which the $SO(2,d-1)$ isometry group transformations act linearly was considered in Ref.~\cite{MN-1}. It is also possible to consider particle and string models \cite{Marnelius}, \cite{Bars} in $(d+2)-$dimensional space with linearly realized $SO(2,d)$ transformations, which describe propagation on either $d-$dimensional Minkowski or anti-de Sitter space depending on the conditions imposed to partially fix extended gauge symmetries. To examine conformal field theories widely used is a realization of Minkowski (Euclidean) space as a projective light-cone in the space-time with two extra dimensions \cite{Salam}, \cite{Ferrara}. Massless (spinning) particle and null string models on the projective light-cone (named conformal space there) were considered in \cite{Martensson}, \cite{GLSSV-conf-string}, \cite{Saltsidis-conf-string}.
More recently projective light-cone (embedding space) approach has been applied to obtain correlation functions in $d$-dimensional conformal field theories taking advantage of the AdS/CFT inspired techniques \cite{CCP}, \cite{Weinberg}, \cite{Penedones}, as well as to study higher-spin field dynamics on $AdS_d$ and its conformal boundary \cite{Grig}, \cite{Didenko}. In \cite{Williams} there was considered the possibility of applying the twistor methods to the $AdS/CFT$ duality relying on the projective-space description of $AdS_{d}$ that naturally combines linear realization of $SO(2,d-1)$ isometry with the projective light-cone description of the $(d-1)$-dimensional conformal boundary.

Utility of the projective-space realization of anti-de Sitter space from the viewpoint of the canonical description of massless particle (tensionless string)  mechanics can be illustrated as follows. Description of $AdS_d$ as an embedded hyperboloid assumes imposition of the constraint $x^2+1\approx0$ on the ambient-space coordinates $x^{\un m}$, where the $SO(2,d-1)$-invariant scalar product $x^2=(x\cdot x)=x^{\un{m}}\eta_{\un m\un n}x^{\un n}$ is taken using Minkowski metric $\eta_{\un m\un n}=\mathrm{diag}(-,+,\ldots,+,-)$ and $AdS_d$ radius is set to unity. In the canonical approach the mass-shell constraint for the massless particle (tensionless string zero modes) in its simplest form is $p^2\approx0$ and its
Poisson bracket (P.B.) relations with the above constraint imply that $(x\cdot p)\approx0$ is also a constraint forming with $x^2+1\approx0$ the pair of the second-class constraints. Quantization in the presence of the second-class constraints necessitates introduction of the Dirac brackets (D.B.) that in general essentially complicates further analysis so it is convenient to consider $x^2+1\approx0$ as a gauge-fixing condition for the constraint $(x\cdot p)\approx0$ that generates dilatations of the embedding-space coordinates \cite{Bonelli'03}. But gauging dilatations precisely implements the projective-space realization of $AdS_d$ so the set of two first-class constraints $(x\cdot p)\approx0$ and $p^2\approx0$ can be taken as the starting point for describing massless particle (tensionless string zero modes) Hamiltonian on $AdS_d$ in such an approach.

In the Lagrangian formalism important feature of the projective-space description of $AdS_d$ is that the object that can be identified with the metric, taking into account the form of the line element,
\beq
ds^2=\frac{1}{|x|^2}dx^{\un m}\theta_{\un m\un n}dx^{\un n},\quad\theta_{\un m\un n}=\eta_{\un m\un n}+\frac{1}{|x|^2}x_{\un m}x_{\un n},
\eeq
is degenerate $\mathrm{det}\,\theta=0$. It can be easily shown to have the eigenvector $x^{\un m}$ with zero eigenvalue $\theta_{\un m\un n}x^{\un n}=0$. So in this approach one is led to consider particle (string, brane) mechanics in the space with degenerate metric. Hamiltonian mechanics of particle and string models in curved spaces with degenerate metrics has been previously treated in \cite{Rivelles}, though the results appear to strongly depend on the particular form of the metric indicating favorability of the case by case study. 
In the higher-spin theory on $AdS_d$ $\theta_{\un m\un n}$ is known as a projection operator that enters field equations in the embedding-space formulation \cite{Fronsdal}, \cite{FangF}, \cite{Metsaev'95}, \cite{Metsaev'97}.

Here we adhere to the bottom-up approach (see, e.g., Ref.~\cite{Brink}): starting with the world-line (world-sheet) supersymmetry generator that is the classical analogue of the Dirac-type equation on $AdS_d$ in the projective-space formulation we seek to close the P.B. (D.B.) algebra of the constraints that includes the generators of the world-line (world-sheet) reparametrizations and space-time dilatations and then proceed to write down corresponding phase-space Lagrangian based on the (weakly vanishing) Hamiltonian. Upon integrating out the space-time momentum and the Lagrange multipliers for the first-class constraints this Lagrangian can further be expressed in terms of the configuration-space variables in various forms.

In the next section the approach outlined above is applied to the spinning particle on the $AdS_d$ space-time in the projective-space formulation. After suggesting closed algebra of the first-class constraints including the generators of $1d$ supersymmetry, world-line reparametrizations and space-time dilatations we discuss various representations of the particle Lagrangian. Further it is considered quantum realization of the classical constraint algebra and obtained are the first-order (Dirac-type) and second-order (Klein-Gordon-type) equations for the wave function of the spinning particle. Section 3 is devoted to the study of the null string on $AdS_d$. Compatibility with the world-sheet conformal reparametrizations essentially restricts the form of admissible constraints. We have found closed classical algebra for the quadratic first-class constraints which include world-sheet supersymmetry, space-time dilatations and Virasoro generators and examine various representations of the associated null string action. Then we discuss two quantum realizations of the constraint algebra: one corresponding to the generalization of the coordinate-momentum ordering and another in terms of positive and negative frequency modes of the operators associated with the classical phase-space variables. The former realization is anomaly free for any dimension of the anti-de Sitter space-time, while for the latter dilatations and conformal reparametrizations are anomalous. Complete values of these anomalies are computed in the framework of the BRST approach.

\setcounter{equation}{0}
\section{Spinning particle}

In what follows we consider $AdS_d$ space-time to be parametrized by $d+1$ homogeneous coordinates $x^{\un m}$ ($\un m=0,\ldots,d$), for which the canonical momenta are $p_{\un m}$. Spinning degrees of freedom of the particle are described by the $SO(2,d-1)$ vector $\xi^{\un m}$ with the Grassmann-odd components.

\subsection{Classical mechanics}

We start with the classical analogue of the Dirac equation
\beq\label{susy-constr}
\Phi=|x|(\xi\cdot p)
\approx0,
\eeq
where $|x|=\sqrt{-x^2}$ and $(\xi\cdot p)=\xi^{\un m}\eta_{\un m\un n}p^{\un n}$ are the $SO(2,d-1)$-invariant norm and scalar product in the embedding space.
With the standard definition of the P.B. and D.B. relations
\beq\label{pointp-pb}
\{p_{\un m},x^{\un n}\}_{P.B.}=\de_{\un m}^{\un n},\quad\{\xi^{\un m},\xi^{\un n}\}_{D.B.}=i\eta^{\un m\un n}
\eeq
it is possible to obtain the closed algebra of three constraints: odd $\Phi$ and bosonic $T$ and $D$
\beq\label{bos-constr}
T=|x|^2p^2+2i(\xi\cdot x)(\xi\cdot p)
\approx0,\quad D=(x\cdot p)\approx0.
\eeq
The only non-zero D.B. relation of this algebra is
\beq\label{1d-susy-db}
\{\Phi,\Phi\}_{D.B.}=iT
\eeq
so that it can be recognized as a $1-$dimensional supersymmetry algebra. In the bosonic limit our constraint algebra reduces to that spanned by $D\approx0$ and $T\approx0$ \cite{Bonelli'03}.

This allows to write the spinning particle action in terms of the phase-space variables
\beq\label{phs-lagrangian}
S=\int d\tau\mathscr L_{\mathrm{ph}},\quad\mathscr L_{\mathrm{ph}}=(p\cdot\dot x)+\frac{i}{2}(\xi\cdot\dot\xi)-\frac{e}{2}T+aD+i\chi\Phi,
\eeq
where the constraints (\ref{susy-constr}) and (\ref{bos-constr}) are introduced with the Lagrange multipliers $e$ and $a$ which are even and $\chi$ which is odd. Integrating out the momentum $p_{\un m}$ yields configuration-space form of the particle's Lagrangian
\beq
\mathscr L_{\mathrm c}=\frac{1}{2e|x|^2}(\dot x+ax)^2+\frac{i}{2}(\xi\cdot\dot\xi)-\frac{i}{|x|^2}(\xi\cdot x)(\xi\cdot\dot x)+\frac{i\chi}{e|x|}\xi\cdot(\dot x+ax).
\eeq
The form of the Lagrangian $\mathscr L_{\mathrm c}$ suggests that the Lagrange multiplier $a$ plays the role of the $1d$ gauge field for the scale transformations of $x$ and $p$. The non-supersymmetric Lagrangian coincides with that of the conformal particle model of Ref.~\cite{MN-1}. Further integrating out $a$ one arrives at the Lagrangian
\beq
\mathscr L_{RP^d}=\frac{1}{2e|x|^2}(\dot x\theta\dot x)
+\frac{i}{2}(\xi\cdot\dot\xi)-\frac{i}{|x|^2}(\xi\cdot x)(\xi\cdot\dot x)+\frac{i\chi}{e|x|}(\xi\theta\dot x),
\eeq
where $\theta^{\un m\un n}=\eta^{\un m\un n}+\frac{1}{|x|^2}x^{\un m}x^{\un n}$ is the degenerate metric tensor corresponding to the realization of $AdS_d$ as the projective space $RP^d$ parametrized by the homogeneous coordinates.

In the gauge $e=1$, $\chi=0$ spinning particle equations
\beq
\begin{array}{rcl}
\frac{d}{d\tau}\left(\frac{1}{|x|}(\theta\dot x)^{\un m}-\frac{i}{|x|}(\xi\cdot x)\xi^{\un m}\right)&-&\frac{1}{|x|^3}(\dot x\theta\dot x)x^{\un m}
-\frac{i}{|x|}(\xi\theta\dot x)\xi^{\un m}\\[0.2cm]
&-&
\frac{2i}{|x|^3}(\xi\cdot\dot x)(\xi\cdot x)x^{\un m}=0,\\[0.3cm]
\dot\xi^{\un m}&+&\frac{1}{|x|^2}(\xi\cdot x)\dot x^{\un m}-\frac{1}{|x|^2}(\xi\cdot\dot x)x^{\un m}=0
\end{array}
\eeq
admit Lax representation
\beq\label{lax-eq}
\dot L_\tau-M_\tau L_\tau=\dot\xi-M_\tau\xi=0
\eeq
with
\beq\label{lax-pair}
\begin{array}{rcl}
L^{\un m}_\tau&=&\frac{1}{|x|}(\theta\dot x)^{\un m}-\frac{i}{|x|}(\xi\cdot x)\xi^{\un m},\\[0.3cm]
M^{\un m\un n}_\tau&=&\frac{1}{|x|^2}(x^{\un m}\dot x^{\un n}-\dot x^{\un m}x^{\un n})-\frac{i}{|x|^2}(\xi\cdot x)(x^{\un m}\xi^{\un n}-\xi^{\un m}x^{\un n})-i\xi^{\un m}\xi^{\un n}.
\end{array}
\eeq
Observe that $M^{\un m\un n}_\tau$ coincide up to the sign with the $SO(2,d-1)$ generators. The first two summands in the gauge $e=1$, $\chi=0$ equal $x^{\un m}p^{\un n}-x^{\un n}p^{\un m}$ and give the orbital part of the $SO(2,d-1)$ generators. The last summand represents the spin part.

\subsection{Quantization}

Quantization consists in replacing classical observables with the
Hermitian operators. Operators associated with the phase-space variables satisfy the (anti)commutation
relations\footnote{Hats are not placed over the operator quantities to
simplify the notation.}
\beq [p_{\un m},x^{\un n}]=-i\de_{\un
m}^{\un n},\quad\{\xi^{\un m},\xi^{\un n}\}=\eta^{\un m\un n}.
\eeq
It is clear that $\xi^{\un m}=2^{-1/2}\g^{\un m}$ and in
what follows we use $\g-$matrices in $(d+1)$ dimensions.
Hermiticity of $\xi^{\un m}$ is understood in the same sense as
that of $\g^{\un m}$, i.e. $(\g^{\un m})^\dagger=(-)^tA\g^{\un
m}A^{-1}$, where $A=\g^{0_1}\g^{0_2}\cdots\g^{0_t}$ and $t$ is the
number of time-like dimensions ($t=2$ for the ambient space of
$AdS_d$). 

Hermitian operator corresponding to the classical constraint $\Phi$ (\ref{susy-constr})
\beq
\Phi\;\rightarrow\;\frac{1}{\sqrt2}\Phi_{\mathrm H},
\eeq
is defined by the expression
\beq\label{qu-susy}
\Phi_{\mathrm H}=|x|(\g\cdot p)+\frac{i}{2|x|}(\g\cdot x)\approx 0,
\eeq
where the last summand comes
from moving the momentum operator to the right in the manifestly
Hermitian expression $\frac12(|x|p_{\un m}+p_{\un m}|x|)$ and the overall factor of $2^{-1/2}$ has been extracted from $\Phi_{\mathrm H}$ to simplify the form of the quantum counterpart of the classical world-line supersymmetry algebra (\ref{1d-susy-db})
\beq\label{q-susy-alg}
\Phi^2_{\mathrm{H}}=T_{\mathrm{H}}.
\eeq
Fulfilment of (\ref{q-susy-alg}) allows to fix unambiguously the form of the Hermitian operator for the bosonic constraint (\ref{bos-constr})\footnote{Further discussion of ambiguities in the
definition of Hermitian operators in supersymmetric models can be
found, e.g., in \cite{Mcfarlane}, \cite{Smilga}.} 
\beq\label{qu-repa}
T_{\mathrm{H}}=|x|^2p^2+i(\g\cdot x)(\g\cdot p)+i(x\cdot
p)
+\frac{2d+1}{4}\approx0.
\eeq
For another bosonic constraint Hermitian operator realization is chosen to be
\beq\label{qu-dilat}
D_{\mathrm{H}}=(x\cdot
p)-\frac{i(d+1)}{2}\approx0.
\eeq

To apply quantum first-class constraints (\ref{qu-susy}), (\ref{qu-repa}) and (\ref{qu-dilat}) to the configuration-space wave function $\Psi(x)$
the momentum operator has to be realized as the differential
operator. For the case of curved configuration space appropriate
expression for the Hermitian momentum operator is known to be
\beq
p_{\un m}=-i(-g)^{-1/4}\partial_{\un m}(-g)^{1/4},
\eeq
where $g$ is
the determinant of the configuration space metric tensor. In the
projective description of anti-de~Sitter space-time scale-invariant measure can be defined as $|x|^{-d-1}\varep_{\un m_1\un
m_2\cdots\un m_{d+1}}x^{\un m_1}d x^{\un m_2}\wedge\cdots\wedge d x^{\un m_{d+1}}$ so as the definition of the Hermitian
momentum operator we take
\beq
p_{\un m}=-i|x|^{\frac{d+1}{2}}\partial_{\un
m}|x|^{-\frac{d+1}{2}}=-i\left(\partial_{\un m}+\frac{(d+1)}{2|x|^2}x_{\un
m}\right).
\eeq
As a result the operator (\ref{qu-susy}) gives the Dirac-type equation
\beq\label{dirac-eq}
\mathscr
D_{\mathrm{H}}\Psi(x)=\left(|x|(\g\cdot\partial)+\frac{d}{2|x|}(\g\cdot
x)\right)\Psi(x)=0,
\eeq
while the quantum bosonic constraints (\ref{qu-repa}) and (\ref{qu-dilat}) in the configuration space
acquire the form
\beq
\begin{array}{l}
T_{\mathrm{H}}=-|x|\partial_{\un m}(|x|\partial^{\un m})+(\g\cdot
x)(\g\cdot\partial)-(d+1)(x\cdot\partial)-\frac{d^2}{4}\approx0,\\[0.2cm]
D_{\mathrm{H}}=-i(x\cdot\partial)\approx0.
\end{array}
\eeq
So that the wave function
$\Psi(x)$ has homogeneity degree zero in $x^{\un m}$ and satisfies
the second-order equation
\beq\label{kg-eq}
\mathscr
D_{\mathrm{H}}^2\Psi(x)=\left(|x|\partial_{\un m}(|x|\partial^{\un
m})-(\g\cdot
x)(\g\cdot\partial)+\frac{d^2}{4}\right)\Psi(x)=0.
\eeq

In terms of non-homogeneous embedding
coordinates $y^{\un m}=x^{\un m}/|x|$ equations (\ref{dirac-eq}) and (\ref{kg-eq}) become
\beq
\begin{array}{l}
\left((\g\cdot\nabla)+\frac{d}{2}(\g\cdot
y)\right)\Psi(y)=0,\\[0.2cm]
\left((\g\cdot\nabla)^2+\frac{d^2}{4}\right)\Psi(y)=0,
\end{array}
\eeq
where $\nabla=\partial_y+y(y\cdot\partial_y)$,
$\partial_y=\partial/\partial y$ and
$D_{\mathrm{H}}$ trivializes.

\setcounter{equation}{0}
\section{Null spinning string}

\subsection{Classical theory}

In case of the null string P.B. relations (\ref{pointp-pb}) generalize to
\beq\label{string-pb}
\{P_{\un m}(\s),X^{\un n}(\s')\}_{P.B.}=\de_{\un m}^{\un n}\de(\s-\s'),\quad\{\Xi^{\un m}(\s),\Xi^{\un n}(\s')\}_{D.B.}=i\eta^{\un m\un n}\de(\s-\s').
\eeq
We concentrate on the closed null string model with the space-like world-sheet coordinate $\s$ ranging from 0 to $2\pi$.
Since extended nature of the null string puts severe restrictions on possible symmetries consider the following set of the quadratic constraints generalizing (\ref{susy-constr}) and (\ref{bos-constr})
\beq\label{susy-str-constr}
\Phi(\s)=(\Xi\cdot P)\approx0
\eeq
and
\beq\label{bose-str-constr}
T(\s)=P^2\approx0,\quad-L(\s)=(P\cdot\pt_\s X)+\frac{i}{2}(\Xi\cdot\pt_\s\Xi)\approx0,\quad D(\s)=(X\cdot P)\approx0
\eeq
that form closed P.B. algebra
\beq\label{pb-constr-string}
\begin{array}{c}
\{\Phi(\s),\Phi(\s')\}_{P.B.}=iT(\s)\de(\s-\s');\\[0.2cm]
\{\Phi(\s),D(\s')\}_{P.B.}=\Phi(\s)\de(\s-\s'),\\[0.2cm]
\{\Phi(\s),L(\s')\}_{P.B.}=\frac32\Phi(\s)\pt_\s\de(\s-\s')+\pt_\s\Phi(\s)\de(\s-\s');\\[0.2cm]
\{T(\s),D(\s')\}_{P.B.}=2T(\s)\de(\s-\s'),\\[0.2cm]
\{T(\s),L(\s')\}_{P.B.}=2T(\s)\pt_\s\de(\s-\s')+\pt_\s T(\s)\de(\s-\s'),\\[0.2cm]
\{D(\s),L(\s')\}_{P.B.}=D(\s)\pt_\s\de(\s-\s')+\pt_\s D(\s)\de(\s-\s'),\\[0.2cm]
\{L(\s),L(\s')\}_{P.B.}=2L(\s)\pt_\s\de(\s-\s')+\pt_\s L(\s)\de(\s-\s').
\end{array}
\eeq
Note that $L(\s)$ can be identified with the Virasoro generator density, $D(\s)$ -- with the dilatation generator density and the algebra (\ref{pb-constr-string}) -- with the semidirect sum of generators of dilatations, $2d$ supersymmetry and Virasoro transformations.

\subsubsection{Various representations of null string Lagrangian}

Now we can write down the null-string Lagrangian and the action in terms of the phase-space variables
\beq\label{null-phs-lagrangian-fixed}
S=\int d\tau d\s\mc L_{\mathrm{ph}}(\tau,\s),\quad\mc L_{\mathrm{ph}}(\tau,\s)=(P\cdot\pt_\tau X)+\frac{i}{2}(\Xi\cdot\pt_\tau\Xi)-\frac{e}{2}T-vL+aD+i\chi\Phi,
\eeq
where $a$, $e$ and $v$ are even Lagrange multipliers for the bosonic constraints and $\chi$ is an odd Lagrange multiplier for the fermionic constraint $\Phi$. Integrating out the momentum $P^{\un m}(\tau,\s)$ gives configuration-space Lagrangian
\beq\label{null-cs-lagrangian-fixed}
\begin{array}{rcl}
\mc L_{\mathrm{c}}&=&\frac{1}{2e}(\pt_\tau X^{\un m}+v\pt_\s X^{\un m}+aX^{\un m})^2+\frac{i}{2}(\Xi\cdot\pt_\tau\Xi)+\frac{iv}{2}(\Xi\cdot\pt_\s\Xi)\\[0.2cm]
&+&\frac{i\chi}{e}\,\Xi_{\un m}(\pt_\tau X^{\un m}+v\pt_\s X^{\un m}+aX^{\un m}).
\end{array}
\eeq
This Lagrangian can be written in the manifestly
$2d-$covariant form
\beq\label{null-conf-lagrangian} \mc
L_{2d\,\mathrm{cov}}=\frac12(\rho^\mu\pt_\mu X^{\un m}+aX^{\un
m})^2+\frac{i}{2}(\Xi\cdot\rho^\mu\pt_\mu\Xi)+i\chi\Xi_{\un
m}(\rho^\mu\pt_\mu X^{\un m}+aX^{\un m})
\eeq
by introducing 
$\rho^\mu=e^{-1/2}(1,v)$, $\mu=(\tau,\s)$ and redefining vector $\Xi^{\un m}$ and the Lagrange multipliers $a$ and $\chi$ 
\beq 
\Xi^{\un m}\rightarrow e^{-1/4}\Xi^{\un m},\quad a\rightarrow e^{1/2}a,\quad\chi\rightarrow e^{3/4}\chi.
\eeq 

Configuration-space form of the null string Lagrangian makes
obvious the role of the Lagrange multiplier $a$ as a gauge field
for the scaling symmetry similarly to the spinning particle model.
We can integrate it out to obtain the Lagrangian
\beq\label{second-order-lagran} \mc
L_{RP^d}=\frac{1}{2}\rho^\mu\rho^\nu(\pt_\mu X\theta\pt_\nu
X)+\frac{i}{2}(\Xi\cdot\rho^\mu\pt_\mu\Xi)+i\chi(\Xi\,\theta\rho^\mu\pt_\mu
X)
\eeq
that manifests projective space realization of the $AdS_d$ space-time and is the generalization of the conformal null
spinning string Lagrangian of Ref.~\cite{Saltsidis-conf-string} (see
also \cite{GLSSV-conf-string}). This Lagrangian can be rewritten
in yet other forms that exhibit dependence on the world-sheet vector density $\rho^\mu$
\beq\label{second-order-lagran2}
\begin{array}{rcl}
\mc L_{RP^d}&=&\frac{1}{2}(\theta^{\un m\un n}\rho^\mu\pt_\mu X_{\un n}+i\chi\Xi^{\un m})^2+\frac{i}{2}(\Xi\cdot\rho^\mu\pt_\mu\Xi)\\[0.2cm]
&=&\frac{1}{2}\rho^\mu\rho^\nu g_{\mu\nu}+\rho^\mu\zeta_\mu,
\end{array}
\eeq
where $g_{\mu\nu}=(\pt_\mu X\theta\pt_\nu X)$ is the induced
world-sheet metric and
$\zeta_{\mu}=\frac{i}{2}(\Xi\cdot\pt_\mu\Xi)+i\chi(\Xi\,\theta\pt_\mu
X)$. It follows that the $SO(2,d-1)$ vector $\theta^{\un m\un
n}\rho^\mu\pt_\mu X_{\un n}+i\chi\Xi^{\un m}$ is null, whereas
equations for $\rho^\mu$
\beq\label{rho-eq-spin}
g_{\mu\nu}\rho^\nu+\zeta_\nu=0
\eeq
imply that induced
world-sheet metric is non-degenerate (see
\cite{LST-null-spin-string}). Thus Eq.~(\ref{rho-eq-spin}) has the
unique solution $\rho^\mu=-g^{-1\,\mu\nu}\zeta_\nu$ and its
substitution back into (\ref{second-order-lagran2}) yields
non-polynomial form of the null spinning string Lagrangian
\beq \mc
L'_{RP^d}=\frac{1}{2g}\zeta_\mu\varepsilon^{\mu\nu}g_{\nu\lambda}\varepsilon^{\lambda\rho}\zeta_{\rho},\quad
g=\mathrm{det}g_{\mu\nu}.
\eeq
This is in contrast with the
bosonic null string for which the equation
\beq\label{rho-eq-bos}
g_{\mu\nu}\rho^\nu=0
\eeq
implies that $g=0$. Solving
(\ref{rho-eq-bos}) under assumption that $g_{\s\s}\not=0$ and
$\rho^\tau\not=0$ and substituting the solution into the bosonic null string Lagrangian
\beq
\mc L_{\mathrm{bos null}}=\rho^\mu\rho^\nu g_{\mu\nu}
\eeq
brings it to the form \beq
\mc L'_{\mathrm{bos
null}}=\frac{g}{E},\quad E=\frac{g_{\s\s}}{(\rho^\tau)^2}
\eeq
which directly generalizes massless particle Lagrangian and for the
case of the null string in flat space-time was examined in \cite{Karlhede},
\cite{AAZ'87}.

\subsubsection{Equations of motion and gauge conditions}

Linear equations for the null-string phase-space variables and the constraints are obtained from the variation of the first-order action (\ref{null-phs-lagrangian-fixed})
\begin{eqnarray}
\dot X^{\un m}+v\acute X^{\un m}+aX^{\un m}+i\chi\Xi^{\un m}&=&eP^{\un m},\nonumber \\[0.2cm]
\dot P^{\un m}+(vP^{\un m})'&=&aP^{\un m},\label{null-1st-order-eom}\\[0.2cm]
\dot\Xi^{\un m}+v\acute\Xi^{\un m}+\frac12\acute v\Xi^{\un m}&=&\chi P^{\un m}.\nonumber
\end{eqnarray}

Above equations simplify if one takes into account gauge symmetries of the action (\ref{null-phs-lagrangian-fixed}) generated by the first-class constraints (\ref{susy-str-constr}) and (\ref{bose-str-constr}). Fermionic constraint (\ref{susy-str-constr}) is responsible for local world-sheet supersymmetry
\beq
\de_\varep X^{\un m}=-\varep\Xi^{\un m},\quad\de_\varep\Xi^{\un m}=-i\varep P^{\un m},\quad\de_\varep e=-2\varep\chi,\quad\de_\varep\chi=-i\left(\dot\varep+v\acute\varep+a\varep-\frac12\acute v\varep\right),
\eeq
where $\varep(\tau,\s)$ is the odd parameter.
Bosonic constraint $D(\s)$ generates local dilations with the parameter $\Delta(\tau,\s)$
\beq
\de_\Delta X^{\un m}=-\Delta X^{\un m},\quad\de_\Delta P^{\un m}=\Delta P^{\un m},\quad\de_\Delta a=\dot\Delta+v\acute\Delta,\quad\de_\Delta e=-2\Delta e,\quad\de_\Delta\chi=-\Delta\chi.
\eeq
$T(\s)$ and $L(\s)$ generate time-like and space-like world-sheet reparametrizations with parameters $\mu(\tau,\s)$ and $\la(\tau,\s)$
\beq
\de_\mu X^{\un m}=-2\mu P^{\un m},\quad\de_\mu e=-2(\dot\mu+v\acute\mu+2a\mu-\acute v\mu)
\eeq
and
\beq
\begin{array}{c}
\de_\la X^{\un m}=\la\acute X^{\un m},\quad\de_\la P^{\un m}=(\la P^{\un m})^\prime,\quad\de_\la\Xi^{\un m}=\la\acute\Xi^{\un m}+\frac12\acute\la\Xi^{\un m},\\[0.2cm]
\de_\la a=\la\acute a,\quad\de_\la e=\la\acute e-\acute\la e,\quad\de_\la v=-\dot\la-\acute\la v+\la\acute v,\quad\de_\la\chi=\la\acute\chi-\frac12\acute\la\chi.
\end{array}
\eeq

Below we consider equations (\ref{null-1st-order-eom}) in a number of widely used gauges. In all these gauges Lagrange multipliers $a$ and $\chi$ are set to zero.
\beq\label{plg1}
\mbox{Gauge 1 (particle-like)}:\quad v=0,\quad e=1.
\eeq
In this gauge e.o.m. reduce to
\beq
\dot X^{\un m}=P^{\un m},\quad\dot P^{\un m}=\dot\Xi^{\un m}=0.
\eeq
Their solutions are
\beq
X^{\un m}(\tau,\s)=X^{\un m}_0(\s)+\tau P^{\un m}_0(\s),\quad P^{\un m}(\tau,\s)=P^{\un m}_0(\s),\quad\Xi^{\un m}(\tau,\s)=\Xi^{\un m}_0(\s),
\eeq
where $X^{\un m}_0(\s)$, $P^{\un m}_0(\s)$ and $\Xi^{\un m}_0(\s)$ represent initial data. For the closed null string $X^{\un m}_0(\s)$ and $P^{\un m}_0(\s)$ are sigma-periodic $\s\simeq\s+2\pi$, while $\Xi^{\un m}_0(\s)$ can also be antiperiodic to account for the Neveu-Schwarz (NS) sector states. In this gauge constraints (\ref{susy-str-constr}) and (\ref{bose-str-constr}) acquire the on-shell form
\beq\label{constr-evolution1}
\begin{array}{c}
T(\tau,\s)=T_0(\s)\approx0,\quad D(\tau,\s)=D_0(\s)+\tau T_0(\s)\approx0,\quad L(\tau,\s)=L_0(\s)-\frac12\tau\acute T_0(\s)\approx0;\\[0.2cm]
\Phi(\tau,\s)=\Phi_0(\s)\approx0,
\end{array}
\eeq
where
\beq\label{initial-constr}
\begin{array}{c}
T_0(\s)=P^2_0\approx0,\quad D_0(\s)=(X_0\cdot P_0)\approx0,\quad-L_0(\s)=(P_0\cdot\acute X_0)+\frac{i}{2}(\Xi_0\cdot\acute\Xi_0)\approx0;\\[0.2cm]
\Phi_0(\s)=(P_0\cdot\Xi_0)\approx0
\end{array}
\eeq
represent initial data on the constraint surface. Eq.~(\ref{constr-evolution1}) illustrates the fact that time evolution does not bring the constraints out of the constraint surface defined by the above initial data.

\beq\label{plg2}
\mbox{Gauge 2 (static)}:\quad v=0,\quad e=0.
\eeq
In this gauge e.o.m. trivialize $\dot X^{\un m}=\dot P^{\un m}=\dot\Xi^{\un m}=0$ so that the solution is defined by the null string initial profile
\beq
X^{\un m}(\tau,\s)=X^{\un m}_0(\s),\quad P^{\un m}(\tau,\s)=P^{\un m}_0(\s),\quad\Xi^{\un m}(\tau,\s)=\Xi^{\un m}_0(\s)
\eeq
and the constraints are  'frozen'
\beq\label{constr-evolution2}
\begin{array}{c}
T(\tau,\s)=T_0(\s)\approx0,\quad D(\tau,\s)=D_0(\s)\approx0,\quad L(\tau,\s)=L_0(\s)\approx0;\\[0.2cm]
\Phi(\tau,\s)=\Phi_0(\s)\approx0
\end{array}
\eeq
with the initial data on the constraint surface defined in (\ref{initial-constr}).

Next consider two gauges for which Lagrange multiplier $v$ is a non-zero constant.
\beq\label{sg3}
\mbox{Gauge 3 (string-like)}:\quad v=-1,\quad e=2.
\eeq
E.o.m. acquire the string-like form
\beq
\pt_-X^{\un m}=P^{\un m},\quad\pt_-P^{\un m}=0,\quad\pt_-\Xi^{\un m}=0,
\eeq
with $\pt_{\pm}=\frac12(\pt_\tau\pm\pt_\s)$, and their solution has the form similar to that in the particle-like gauge
\beq
X^{\un m}(\tau,\s)=X^{\un m}_0(\s^+)+\s^- P^{\un m}_0(\s^+),\quad P^{\un m}(\tau,\s)=P^{\un m}_0(\s^+),\quad\Xi^{\un m}(\tau,\s)=\Xi^{\un m}_0(\s^+),
\eeq
where $\s^{-}=\tau-\s$ plays the role of the world-sheet time-like variable and $\s^{+}=\tau+\s$ -- of the space-like variable. Superficially in this gauge $\sigma$-periodicity is lost but looking at the constraints, whose modes in quantum theory are used to identify physical subspace of the state space, it is clear that they are weakly periodic
\beq\label{constr-evolution3}
\begin{array}{c}
T(\tau,\s)=T_0(\s^+)\approx0,\quad D(\tau,\s)=D_0(\s^+)+\s^- T_0(\s^+)\approx0,\\[0.2cm]
L(\tau,\s)=L_0(\s^+)+T_0(\s^+)-\frac12\s^-(T_0)^\prime\approx0;\\[0.2cm]
\Phi(\tau,\s)=\Phi_0(\s^+)\approx0.
\end{array}
\eeq
In (\ref{constr-evolution3})
\beq\label{initial-constr3}
\begin{array}{c}
T_0(\s^+)=P^2_0\approx0,\quad D_0(\s^+)=(X_0\cdot P_0)\approx0,\quad-L_0(\s^+)=(P_0\cdot\acute X_0)+\frac{i}{2}(\Xi_0\cdot\acute\Xi_0)\approx0;\\[0.2cm]
\Phi_0(\s^+)=(P_0\cdot\Xi_0)\approx0
\end{array}
\eeq
define initial data on the constraint surface similarly to (\ref{initial-constr}) and ${}^\prime$ stands for the differentiation w.r.t. the function's argument.

Yet another gauge is
\beq\label{cftg4}
\mbox{Gauge 4 (CFT-like)}:\quad v=-1,\quad e=0.
\eeq
In this gauge e.o.m. yield that
\beq
X^{\un m}(\tau,\s)=X^{\un m}_0(\s^+),\quad P^{\un m}(\tau,\s)=P^{\un m}_0(\s^+),\quad\Xi^{\un m}(\tau,\s)=\Xi^{\un m}_0(\s^+)
\eeq
in analogy with the family of the world-sheet $\bt\g$ CFT's with the action (see, e.g., \cite{Pol-book})
\beq
S_{\bt\g}=\int d\tau d\s\bt\pt_-\g
\eeq
and $\psi$ CFT defined by
\beq
S_\psi=i\int d\tau d\s\psi\pt_-\psi.
\eeq
Constraints are defined by their initial values (\ref{initial-constr3}), in particular $L_0$ has the form of the sum of the stress-tensor of $\bt\g$ CFT with conformal weights $w_\bt=1$, $w_\g=0$ and $\psi$ CFT with $w_\psi=1/2$.

As a concluding remark of this subsection note that the null-string action corresponding to the second-order Lagrangian (\ref{second-order-lagran}) is a convenient framework for discussing integrability of the e.o.m. In the gauge $\rho^\tau=1$, $\rho^\s=\chi=0$ equations for the space-time coordinates admit Lax representation similar to that of the spinning particle (\ref{lax-eq}), (\ref{lax-pair})
\beq
\frac{d}{d\tau}\mathbb L_\tau-\mathbb M_\tau\mathbb L_\tau=0:\quad\mathbb L^{\un m}_\tau=\frac{1}{|X|}(\theta\dot X)^{\un m},\quad
\mathbb M^{\un m\un n}_\tau=\frac{1}{|X|^2}(X^{\un m}\dot X^{\un n}-\dot X^{\un m}X^{\un n}).
\eeq

\subsubsection{Fourier expansions of the phase-space variables and constraints}

In all of the above gauges initial data for the phase-space
variables are taken $\s-$periodic and can be Fourier
expanded\footnote{If it is not specified explicitly all sums are
over integers ranging from minus to plus infinity.}
\beq\label{matter-fourier}
\begin{array}{c}
P^{\un m}_0(\s)=\frac{i}{\sqrt{2\pi}}\sum\limits_m\mathrm{e}^{im\s}p^{\un m}_m,\quad
X^{\un m}_0(\s)=\frac{1}{\sqrt{2\pi}}\sum\limits_m\mathrm{e}^{im\s}x^{\un m}_m,\\[0.3cm]
\Xi^{\un m}_0(\s)=\frac{1}{\sqrt{2\pi}}\sum\limits_{m\in\mathbb{Z}+\nu}\mathrm{e}^{im\s}\xi^{\un m}_m.
\end{array}
\eeq
Fermions are allowed to be either periodic ($\nu=0$, Ramond (R) sector) or antiperiodic ($\nu=1/2$, NS sector).
Then one can define Fourier modes of the initial data for the constraints. For the Virasoro generator density
\beq\label{Vir-ini-dens}
L_0(\s)=-(P_0\cdot\acute X_0)-\frac{i}{2}(\Xi_0\cdot\acute\Xi_0)\approx0
\eeq
we define
\beq\label{vir-fourier}
L_0(\s)=\frac{1}{2\pi}\sum\limits_m\mathrm{e}^{im\s}L^{\mathrm{m}}_m,\quad L^{\mathrm{m}}_m=L^{\mathrm{bos}}_m+L^{\mathrm{ferm}}_m,
\eeq
where the superscript $\mathrm{m}$ labels modes of the matter part of the Virasoro generator that in the next section will be augmented by the contributions of ghosts. $L^{\mathrm{bos}}_m$ and $L^{\mathrm{ferm}}_m$ read
\beq\label{Vir-fourier-bos-ferm}
L^{\mathrm{bos}}_m=\sum\limits_nn(p_{m-n}\cdot x_n),\quad L^{\mathrm{ferm}}_m=\frac14\sum\limits_{n\in\mathbb{Z}+\nu}(2n-m)(\xi_{m-n}\cdot \xi_n).
\eeq
Similarly for the dilatation generator density
\beq\label{dil-ini-dens}
D_0(\s)=(X_0\cdot P_0)\approx0
\eeq
we find
\beq\label{dil-fourier}
D_0(\s)=\frac{1}{2\pi}\sum\limits_m\mathrm{e}^{im\s}D^{\mathrm{m}}_m,\quad D^{\mathrm{m}}_m=i\sum\limits_n(p_{m-n}\cdot x_n).
\eeq
For the bosonic constraint
\beq\label{mass-ini-dens}
T_0(\s)=P^2_0\approx0
\eeq
Fourier-mode expansion reads
\beq\label{mass-fourier}
T_0(\s)=\frac{1}{2\pi}\sum\limits_m\mathrm{e}^{im\s}T^{\mathrm{m}}_m,\quad T^{\mathrm{m}}_m=-\sum\limits_n(p_{m-n}\cdot p_n).
\eeq
Finally for the world-sheet supersymmetry generator density
\beq\label{susy-ini-dens}
\Phi_0(\s)=(P_0\cdot\Xi_0)\approx0
\eeq
we have
\beq\label{susy-fourier}
\Phi_0(\s)=\frac{1}{2\pi}\sum\limits_{m\in\mathbb{Z}+\nu}\mathrm{e}^{im\s}\Phi^{\mathrm{m}}_m,\quad\Phi^{\mathrm{m}}_m=i\sum\limits_{n\in\mathbb{Z}+\nu}(p_{m-n}\cdot\xi_n).
\eeq

\subsection{Quantum theory}

Upon quantization P.B. relations (\ref{string-pb}) become (anti)commutators
\beq
[P_{0\,\un m}(\s),X_0^{\un n}(\s')]=-i\de_{\un m}^{\un n}\de(\s-\s'),\quad\{\Xi_0^{\un m}(\s),\Xi_0^{\un n}(\s')\}=\eta^{\un m\un n}\de(\s-\s')
\eeq
which for the modes translate into the relations
\beq
[p_{\un mm},x_n^{\un n}]=-\de_{\un m}^{\un n}\de_{m,-n},\quad\{\xi^{\un m}_m,\xi^{\un n}_n\}=\eta^{\un m\un n}\de_{m,-n}.
\eeq
Since operators $P_{0\un m}(\s)$, $X_0^{\un m}(\s)$ and $\Xi_0^{\un m}(\s)$ are assumed to be Hermitian their modes have the following conjugation properties
\beq
(p_m^{\un m})^\dagger=-p_{-m}^{\un m},\quad (x_m^{\un m})^\dagger=x_{-m}^{\un m},\quad(\xi^{\un m}_m)^\dagger=\xi^{\un m}_{-m}.
\eeq

In quantum theory P.B. relations of the classical null string constraint
algebra (\ref{pb-constr-string}) may acquire anomalous
contributions depending on the choice of the vacuum and ordering
of the constituent Fourier modes (\ref{matter-fourier}). Below we
discuss two vacua commonly used in the quantization of
null strings: one that is annihilated by the  momentum conjugate
to the space-time coordinates \cite{BZh}, \cite{ILST'93} and
corresponds to the coordinate-momentum or $xp$-ordering, and
another -- annihilated by the positive-frequency modes of
coordinates and momenta \cite{GRRA} that corresponds to placing
positive-frequency modes to the right of the negative-frequency ones that is to the $(-+)$-ordering.\footnote{For the
recent comparative analysis of those vacua in the context of
ambitwistor strings \cite{Mason'13} see \cite{Tourkine}. The relationship between
(ambi)twistor strings and the tensionless (limit of) string was
also discussed in \cite{Bandos'14} (see also \cite{BdAM}).} For
each choice of the vacuum for the bosonic variables there is a
natural associated vacuum for the fermions and ghosts. Potentially
anomalous are (anti)commutation relations between operators for
which, using the parlance of the Wick theorem, multiple
cross-contractions of the non-commuting constituent operators are
possible, in our case of the coordinates and momenta and/or
fermionic operators between themselves. Since all the constraints of the
null spinning string are quadratic possible
double contractions may lead to anomalous contributions.
Convenient way of identifying such terms is to consider commutators
of the corresponding Fourier modes of the constraints
(\ref{Vir-fourier-bos-ferm}), (\ref{dil-fourier}),
(\ref{mass-fourier}) and (\ref{susy-fourier}). By inspection
potentially anomalous appear to be the commutators
\beq\label{potential-anomalies}
[L^{\mathrm{m}}_m,L^{\mathrm{m}}_{-m}],\quad
[D^{\mathrm{m}}_m,D^{\mathrm{m}}_{-m}],\quad [D^{\mathrm{m}}_{\pm
m},L^{\mathrm{m}}_{\mp m}].
\eeq
To calculate them it
is convenient to present Fourier modes of involved generators in
the form, in which contributions of the positive and negative modes are written down
explicitly.

\subsubsection{Quantum algebra of the constraints}

Consider in detail the contribution of the bosonic modes in (\ref{Vir-fourier-bos-ferm}) to the first commutator in (\ref{potential-anomalies}). For positive and negative modes of $L^{\mathrm{bos}}_m$ we get
\beq\label{Vir-pos-neg-fourier-modes}
\begin{array}{rl}
L^{\mathrm{bos}}_{m>0}=&\sum\limits_{n=1}^mn(x_n\cdot p_{m-n})+\sum\limits_{n=1}^\infty(m+n)(x_{m+n}\cdot p_{-n})-\sum\limits_{n=1}^\infty n(x_{-n}\cdot p_{m+n}),\\[0.2cm]
L^{\mathrm{bos}}_{-m}=&-\sum\limits_{n=1}^mn(x_{-n}\cdot p_{-(m-n)})-\sum\limits_{n=1}^\infty(m+n)(x_{-(m+n)}\cdot p_{n})+\sum\limits_{n=1}^\infty n(x_{n}\cdot p_{-(m+n)}).
\end{array}
\eeq 
These are Hermitian conjugate
$(L^{\mathrm{bos}}_{m})^\dagger=L^{\mathrm{bos}}_{-m}$ and are free
from the ordering ambiguities unlike $L^{\mathrm{bos}}_{0}$,
Hermitian expression for which will emerge as an outcome of the
commutator calculation. Ordering violation arises only from
commuting finite sums in (\ref{Vir-pos-neg-fourier-modes}) for the
$(-+)$-ordering  leading to the anomalous contribution
\beq
[L^{\mathrm{bos}}_m,L^{\mathrm{bos}}_{-m}]|_{\textrm{$(-+)$-ordering}}=\frac{(d+1)}{6}\,
m^3+2mL^{\mathrm{bos}}_{0(-+)}
\eeq
and $L^{\mathrm{bos}}_{0(-+)}$ is
defined by the following expression
\beq\label{vir-bos-mat-zero-mode-frord}
L^{\mathrm{bos}}_{0(-+)}=\sum\limits_{n=1}^\infty
n\left((p_{-n}\cdot x_{n})-(x_{-n}\cdot
p_{n})\right)-\frac{(d+1)}{12}
\eeq
which is manifestly Hermitian
with the freedom of adding a real constant $-\frac{(d+1)}{12}$ that has been chosen
in accordance with the conventional string-theoretic value for $2(d+1)$ periodic bosons (see, e.g., \cite{Pol-book}). In contrast, for the $xp$-ordering above commutator takes the same form as
in the classical theory
\beq
[L^{\mathrm{bos}}_m,L^{\mathrm{bos}}_{-m}]|_{\textrm{$xp$-ordering}}=2mL^{\mathrm{bos}}_{0\,
xp},
\eeq
where
\beq\label{vir0-mat-bos-yp}
L^{\mathrm{bos}}_{0\,
xp}=\sum\limits_{n=1}^\infty n\left((x_{n}\cdot
p_{-n})-(x_{-n}\cdot p_{n})\right).
\eeq
Checking its Hermiticity
requires Fourier modes to be commuted producing terms proportional to
the infinite sum $\sum\limits_{n=1}^\infty n$ that can be given
finite value $-1/12$ via the $\zeta-$function
regularization conventionally used in string theory. In the final expression these terms cancel out.

Similarly can be calculated other two commutators in (\ref{potential-anomalies}). Making explicit contributions of positive and negative modes in $D^{\mathrm{m}}_m$
\beq\label{dil-mat-modes}
\begin{array}{rl}
D^{\mathrm{m}}_{m>0}=&i\left(\sum\limits_{n=0}^m(x_n\cdot p_{m-n})+\sum\limits_{n=1}^\infty(x_{m+n}\cdot p_{-n})+\sum\limits_{n=1}^\infty(x_{-n}\cdot p_{m+n})\right),\\[0.2cm]
D^{\mathrm{m}}_{-m}=&i\left(\sum\limits_{n=0}^m(x_{-n}\cdot p_{-(m-n)})+\sum\limits_{n=1}^\infty(x_{-(m+n)}\cdot p_{n})+\sum\limits_{n=1}^\infty(x_{n}\cdot p_{-(m+n)})\right)
\end{array}
\eeq we find\footnote{Observe that
$(D^{\mathrm{m}}_{m})^\dagger=D^{\mathrm{m}}_{-m}$.}
\beq
[D^{\mathrm{m}}_m,D^{\mathrm{m}}_{-m}]|_{\textrm{$(-+)$-ordering}}=(d+1)m
\eeq
and
\beq
[D^{\mathrm{m}}_m,D^{\mathrm{m}}_{-m}]|_{\textrm{$xp$-ordering}}=0.
\eeq
Commuting further, e.g., $D^{\mathrm{m}}_{m>0}$ with
$L^{\mathrm{bos}}_{-m}$ gives
\beq
[D^{\mathrm{m}}_{m},L^{\mathrm{bos}}_{-m}]|_{\textrm{$(-+)$-ordering}}=\frac{i(d+1)}{2}m^2+mD^{\mathrm{m}}_{0\,(-+)},
\eeq
where
\beq\label{dil-mat-zero-mode-frord}
D^{\mathrm{m}}_{0\,(-+)}=i\left((x_0\cdot
p_0)+\sum\limits_{n=1}^\infty\left((x_{-n}\cdot p_n)+(p_{-n}\cdot
x_n)\right)-\frac{(d+1)}{2}\right)
\eeq
is the Hermitian zero-mode
operator. For the $xp$-ordering no anomalous terms arise
\beq
[D^{\mathrm{m}}_{m},L^{\mathrm{bos}}_{-m}]|_{\textrm{$xp$-ordering}}=mD^{\mathrm{m}}_{0\,
xp}
\eeq
and $D^{\mathrm{m}}_{0\, xp}$ equals
\beq\label{dil0-mat-yp}
D^{\mathrm{m}}_{0\, xp}=i\left((x_0\cdot
p_0)+\sum\limits_{n=1}^\infty\left((x_{-n}\cdot p_n)+(x_{n}\cdot
p_{-n})\right)\right).
\eeq

Now let us find the contribution to the first commutator in (\ref{potential-anomalies}) of the part of the Virasoro generators (\ref{vir-fourier}) determined by the fermionic modes (\ref{Vir-fourier-bos-ferm}). Precise form of the commutator depends on the periodicity of the fermions. So consider first the Ramond sector. Explicit form of $L^{\mathrm{ferm}}_m$ in terms of the positive and negative frequency modes is
\beq\label{Vir-fermr-pos-neg-fourier-modes}
\begin{array}{rl}
L^{\mathrm{ferm\, R}}_{m>0}=&\frac14\sum\limits_{n=0}^{m}(2n-m)(\xi_{m-n}\cdot\xi_n)-\frac12\sum\limits_{n=1}^\infty(m+2n)(\xi_{m+n}\cdot\xi_{-n}),\\[0.2cm]
L^{\mathrm{ferm\, R}}_{-m}=&\frac14\sum\limits_{n=0}^{m}(m-2n)(\xi_{-(m-n)}\cdot\xi_{-n})+\frac12\sum\limits_{n=1}^\infty(m+2n)(\xi_{-(m+n)}\cdot\xi_{n}).
\end{array}
\eeq
Like for bosons their commutator is anomalous for the
$(-+)-$ordering
\beq\label{Vir-comm-ferm-R-+} [L^{\mathrm{ferm\,
R}}_m,L^{\mathrm{ferm\,
R}}_{-m}]|_{\textrm{$(-+)$-ordering}}=\frac{(d+1)}{24}m^3+2mL^{\mathrm{ferm\,
R}}_{0(-+)}
\eeq
with the Hermitian zero-mode operator
\beq\label{vir-mat-fermr-zero-mode-frord}
L^{\mathrm{ferm\,
R}}_{0(-+)}=\sum\limits_{n=1}^\infty
n(\xi_{-n}\cdot\xi_n)+\frac{(d+1)}{24},
\eeq
where the numerical
constant $\frac{(d+1)}{24}$
is that prescribed for $(d+1)$
periodic fermions \cite{Pol-book}. In contrast there is no
anomaly if the Weyl ordering is chosen for $\xi$-modes in
$L^{\mathrm{ferm\, R}}_0$
\begin{eqnarray}
[L^{\mathrm{ferm\, R}}_m,L^{\mathrm{ferm\, R}}_{-m}]|_{\textrm{Weyl-ordering}}&=&2mL^{\mathrm{ferm\, R}}_{0\, W},\label{Vir-comm-ferm-R-Weyl}\\[0.2cm]
L^{\mathrm{ferm\, R}}_{0\, W}&=&\frac12\sum\limits_{n=1}^\infty n\left((\xi_{-n}\cdot\xi_n)-(\xi_{n}\cdot\xi_{-n})\right). \label{vir0-fermr-weyl}
\end{eqnarray}
Analogous conclusions hold also in the NS sector. $\xi$-modes here are half-integer-valued so that Fourier modes of the Virasoro generator acquire the form
\beq\label{Vir-fermns-pos-neg-fourier-modes}
\begin{array}{rl}
L^{\mathrm{ferm\, NS}}_{m>0}=&\frac14\sum\limits_{r=1/2}^{m-1/2}(2r-m)(\xi_{m-r}\cdot\xi_r)-\frac12\sum\limits_{r=1/2}^\infty(m+2r)(\xi_{m+r}\cdot\xi_{-r}),\\[0.2cm]
L^{\mathrm{ferm\, NS}}_{-m}=&\frac14\sum\limits_{r=1/2}^{m-1/2}(m-2r)(\xi_{-(m-r)}\cdot\xi_{-r})+\frac12\sum\limits_{r=1/2}^\infty(m+2r)(\xi_{-(m+r)}\cdot\xi_{r}).
\end{array}
\eeq
The value of the commutator of $L^{\mathrm{ferm\, NS}}_{m>0}$ and $L^{\mathrm{ferm\, NS}}_{-m}$ for the $(-+)-$ordering is the same as in (\ref{Vir-comm-ferm-R-+})
\beq
[L^{\mathrm{ferm\, NS}}_m,L^{\mathrm{ferm\, NS}}_{-m}]|_{\textrm{$(-+)$-ordering}}=\frac{(d+1)}{24}m^3+2mL^{\mathrm{ferm\, NS}}_{0(-+)}
\eeq
with $L^{\mathrm{ferm\, NS}}_{0(-+)}$ defined by
\beq\label{vir-mat-fermns-zero-mode-frord}
L^{\mathrm{ferm\, NS}}_{0(-+)}=\sum\limits_{r=1/2}^\infty r(\xi_{-r}\cdot\xi_r)-\frac{(d+1)}{48},
\eeq
where each antiperiodic fermion contributes $-1/48$ to the $c-$number term. For the Weyl-ordering similarly to (\ref{Vir-comm-ferm-R-Weyl}) we find
\beq
[L^{\mathrm{ferm\, NS}}_m,L^{\mathrm{ferm\, NS}}_{-m}]|_{\textrm{Weyl-ordering}}=2mL^{\mathrm{ferm\, NS}}_{0\, W}
\eeq
and
\beq\label{vir0-fermns-weyl}
L^{\mathrm{ferm\, NS}}_{0\, W}=\frac12\sum\limits_{r=1/2}^\infty r\left((\xi_{-r}\cdot\xi_r)-(\xi_r\cdot\xi_{-r})\right).
\eeq

So we come to the preliminary conclusion that no anomalies are
observed in the matter sector of the quantum algebra of the constraints
for $xp-$ordering (Weyl-ordering for fermions), while for the
$(-+)-$ordering values of the anomalous contributions are the same
as for the corresponding world-sheet CFTs \cite{Pol-book}.
Complete calculation of the anomalies and study of the possibility of their
cancelation requires also the contributions of
ghosts to be taken into account and should be carried out in the framework of the BRST
quantization to which we now turn.

\subsubsection{BRST quantization}

With each of the constraints (\ref{Vir-ini-dens}), (\ref{dil-ini-dens}), (\ref{mass-ini-dens}) and (\ref{susy-ini-dens}) a canonical pair of ghost and antighost fields is associated that form the triads
\beq
(L_0(\s),c^L(\s),b^L(\s)),\quad(D_0(\s),c^D(\s),b^D(\s)),\quad(T_0(\s),c^T(\s),b^T(\s)),\quad(\Phi_0(\s),\g(\s),\bt(\s)).
\eeq
Classically ghosts satisfy the P.B. relations
\beq
\begin{array}{c}
\{c^L(\s),b^L(\s')\}_{P.B.}=\{c^D(\s),b^D(\s')\}_{P.B.}=\{c^T(\s),b^T(\s')\}_{P.B.}=i\de(\s-\s');\\[0.2cm]
\{\g(\s),\bt(\s')\}_{P.B.}=-\de(\s-\s').
\end{array}
\eeq
Ghosts $\g(\s)$ and $\bt(\s)$ associated with the world-sheet supersymmetry generator $\Phi_0(\s)$ are even, while other ghosts associated with the bosonic constraints are odd, as is the BRST charge
\beq\label{clas-brst-charge}
\Omega=\int\! d\s\, \Omega(\s),\quad\Omega(\s)=c^LL^{\mathrm{ext}}+c^DD^{\mathrm{ext}}+c^TT_0+\g\Phi^{\mathrm{ext}}
\eeq
such that $\{\Omega,\Omega\}_{P.B.}=0$. The density of the BRST charge in (\ref{clas-brst-charge}) has been presented in convenient form as the sum of the constraints (\ref{Vir-ini-dens}), (\ref{dil-ini-dens}), (\ref{mass-ini-dens}), (\ref{susy-ini-dens}) extended by the ghost contributions and multiplied by the associated ghost fields
\beq\label{Vir-constr-ext}
\begin{array}{c}
L^{\mathrm{ext}}(\s)=L_0+\frac12L^{\mathrm{gh\, L}}+L^{\mathrm{gh\, T}}+L^{\mathrm{gh\, D}}+L^{\mathrm{gh}\,\Phi}:\\[0.2cm]
L^{\mathrm{gh\, L(T)}}(\s)=2i(c^{L(T)}b^{L(T)})'-ic^{L(T)}\acute b^{L(T)},\quad L^{\mathrm{gh\, D}}(\s)=i\acute c^Db^D,\quad L^{\mathrm{gh}\,\Phi}(\s)=\g\acute\bt-\frac32(\g\bt)'
\end{array}
\eeq
and
\beq\label{dil-constr-ext}
D^{\mathrm{ext}}(\s)=D_0-2ic^Tb^T+\g\bt,
\eeq
as well as
\beq
\Phi^{\mathrm{ext}}(\s)=\Phi_0-\frac12\g b^T.
\eeq
Observe certain arbitrariness in the definition of extended gauge generators.

By appropriate choice of the gauge fermion, for which P.B. relations with
the BRST charge give the BRST Hamiltonian, it is possible to make
ghost fields satisfy the same equations as the matter fields do
for each of the considered gauge conditions (\ref{plg1}),
(\ref{plg2}), (\ref{sg3}) and (\ref{cftg4}). Besides that they
have the same periodicity as the associated constraints and admit
Fourier expansions similar to those of the matter fields
(\ref{matter-fourier}) \beq
\begin{array}{c}
c(\s)=\frac{1}{\sqrt{2\pi}}\sum\limits_m\mathrm{e}^{im\s}c_m,\quad b(\s)=\frac{1}{\sqrt{2\pi}}\sum\limits_m\mathrm{e}^{im\s}b_m,\\[0.2cm]
\g(\s)=\frac{1}{\sqrt{2\pi}}\sum\limits_{m\in\mathbb{Z}+\nu}\mathrm{e}^{im\s}\g_m,\quad\bt(\s)=\frac{i}{\sqrt{2\pi}}\sum\limits_{m\in\mathbb{Z}+\nu}\mathrm{e}^{im\s}\bt_m,
\end{array}
\eeq
where $c$ and $b$ stand for any of the odd ghosts and both R and NS periodicity conditions for even ghost fields $\g$ and $\bt$ are taken into account.  This enables one to write the BRST charge in terms of modes
\beq
\begin{array}{rl}
\Omega=&c^L_0L^{\mathrm{ext}}_0+\sum\limits_{m=1}^{\infty}(c^L_mL^{\mathrm{ext}}_{-m}+c^L_{-m}L^{\mathrm{ext}}_m)
+c^D_0D^{\mathrm{ext}}_0+\sum\limits_{m=1}^{\infty}(c^D_mD^{\mathrm{ext}}_{-m}+c^D_{-m}D^{\mathrm{ext}}_m)\\[0.2cm]
+&c^T_0T^{\mathrm{m}}_0+\sum\limits_{m=1}^{\infty}(c^T_mT^{\mathrm{m}}_{-m}+c^T_{-m}T^{\mathrm{m}}_m)
+\de_{\nu,0}\g_0\Phi^{\mathrm{ext}}_0+\sum\limits_{m=1-\nu}^{\infty}(\g_m\Phi^{\mathrm{ext}}_{-m}+\g_{-m}\Phi^{\mathrm{ext}}_m).
\end{array}
\eeq

Fourier modes of the extended constraints are given by the sums of contributions of the matter modes and ghosts.
For instance, modes of the extended Virasoro generator (\ref{Vir-constr-ext}) can be presented in the form
\beq\label{vir-ext-modes}
L^{\mathrm{ext}}_m=L^{\mathrm{m}}_m+\frac12L^{\mathrm{gh\, L}}_m+L^{\mathrm{gh\, T}}_m+L^{\mathrm{gh\, D}}_m+L^{\mathrm{gh}\,\Phi}_m.
\eeq
For the matter part (\ref{vir-fourier}) expressions in terms of positive and negative frequency modes have already been found in (\ref{Vir-pos-neg-fourier-modes}), (\ref{Vir-fermr-pos-neg-fourier-modes}) and (\ref{Vir-fermns-pos-neg-fourier-modes}). Contributions of $(c^T,b^T)$ and $(c^D,b^D)$ ghost pairs have the form
\beq\label{mass-ghosts-modes}
\begin{array}{rl}
L^{\mathrm{gh\, T}}_m=&-\sum\limits_{n=0}^m(m+n)c^T_nb^T_{m-n}-\sum\limits_{n=1}^\infty(2m+n)c^T_{m+n}b^T_{-n}-\sum\limits_{n=1}^\infty(m-n)c^T_{-n}b^T_{m+n},\\[0.2cm]
L^{\mathrm{gh\, T}}_{-m}=&\sum\limits_{n=0}^m(m+n)c^T_{-n}b^T_{-(m-n)}+\sum\limits_{n=1}^\infty(2m+n)c^T_{-(m+n)}b^T_n+\sum\limits_{n=1}^\infty(m-n)c^T_nb^T_{-(m+n)}
\end{array}
\eeq
and
\beq\label{dil-ghosts-modes}
\begin{array}{rl}
L^{\mathrm{gh\, D}}_{m}=&-\sum\limits_{n=1}^mnc^D_nb^D_{m-n}-\sum\limits_{n=1}^\infty(m+n)c^D_{m+n}b^D_{-n}+\sum\limits_{n=1}^\infty nc^D_{-n}b^D_{m+n},\\[0.2cm]
L^{\mathrm{gh\, D}}_{-m}=&\sum\limits_{n=1}^mnc^D_{-n}b^D_{-(m-n)}+\sum\limits_{n=1}^\infty(m+n)c^D_{-(m+n)}b^D_n-\sum\limits_{n=1}^\infty nc^D_nb^D_{-(m+n)}.
\end{array}
\eeq
For the Virasoro ghosts $(c^L,b^L)$ the contribution of $\frac12L^{\mathrm{gh\, L}}_m$ is the same as that of $(c^D,b^D)$ ghosts above since $L^{\mathrm{ext}}_m$ is multiplied by $c^L_{-m}$ in the BRST charge. For the bosonic ghosts modes of the Virasoro generator in each sector are
\beq\label{susy-ghosts-modesr}
\begin{array}{rl}
L^{\mathrm{gh}\,\Phi\, \mathrm{R}}_m=&\sum\limits_{n=0}^m(\frac12m+n)\g_n\bt_{m-n}+\sum\limits_{n=1}^\infty(\frac32m+n)\g_{m+n}\bt_{-n}+\sum\limits_{n=1}^\infty(\frac12m-n)\g_{-n}\bt_{m+n},\\[0.2cm]
L^{\mathrm{gh}\,\Phi\, \mathrm{R}}_{-m}=&-\sum\limits_{n=0}^m(\frac12m+n)\g_{-n}\bt_{-(m-n)}-\sum\limits_{n=1}^\infty(\frac32m+n)\g_{-(m+n)}\bt_n-\sum\limits_{n=1}^\infty(\frac12m-n)\g_n\bt_{-(m+n)}
\end{array}
\eeq
and
\beq\label{susy-ghosts-modesns}
\begin{array}{rl}
L^{\mathrm{gh}\,\Phi\, \mathrm{NS}}_m=&\sum\limits_{r=1/2}^{m-1/2}(\frac12m+r)\g_r\bt_{m-r}+\sum\limits_{r=1/2}^\infty(\frac32m+r)\g_{m+r}\bt_{-r}+\sum\limits_{r=1/2}^\infty(\frac12m-r)\g_{-r}\bt_{m+r},\\[0.2cm]
L^{\mathrm{gh}\,\Phi\, \mathrm{NS}}_{-m}=&-\sum\limits_{r=1/2}^{m-1/2}(\frac12m+r)\g_{-r}\bt_{-(m-r)}-\sum\limits_{r=1/2}^\infty(\frac32m+r)\g_{-(m+r)}\bt_r-\sum\limits_{r=1/2}^\infty(\frac12m-r)\g_r\bt_{-(m+r)}.
\end{array}
\eeq

Similarly Fourier modes of the extended dilatation generator
\beq
D^{\mathrm{ext}}_m=D^{\mathrm m}_m+D^{\mathrm{gh\, T}}_m+D^{\mathrm{gh}\,\Phi}_m
\eeq
include contributions of matter variables (\ref{dil-mat-modes}) and ghosts. For the fermionic $(c^T,b^T)$ ghosts we obtain
\beq\label{dil-gen-gh-modes}
\begin{array}{rl}
D^{\mathrm{gh\, T}}_m=&-2i\left(\sum\limits_{n=0}^mc^T_nb^T_{m-n}+\sum\limits_{n=1}^\infty c^T_{m+n}b^T_{-n}+\sum\limits_{n=1}^\infty c^T_{-n}b^T_{m+n}\right),\\[0.2cm]
D^{\mathrm{gh\, T}}_{-m}=&-2i\left(\sum\limits_{n=0}^mc^T_{-n}b^T_{-(m-n)}+\sum\limits_{n=1}^\infty c^T_{-(m+n)}b^T_{n}+\sum\limits_{n=1}^\infty c^T_nb^T_{-(m+n)}\right).
\end{array}
\eeq
For the $(\g,\bt)$ ghosts in R and NS sectors one finds
\beq\label{dil-gen-gh-modes-bosr}
\begin{array}{rl}
D^{\mathrm{gh}\,\Phi\,\mathrm{R}}_m=&i\left(\sum\limits_{n=0}^m\g_n\bt_{m-n}+\sum\limits_{n=1}^\infty \g_{m+n}\bt_{-n}+\sum\limits_{n=1}^\infty \g_{-n}\bt_{m+n}\right),\\[0.2cm]
D^{\mathrm{gh}\,\Phi\,\mathrm{R}}_{-m}=&i\left(\sum\limits_{n=0}^m\g_{-n}\bt_{-(m-n)}+\sum\limits_{n=1}^\infty\g_{-(m+n)}\bt_{n}+\sum\limits_{n=1}^\infty\g_n\bt_{-(m+n)}\right)
\end{array}
\eeq
and
\beq\label{dil-gen-gh-modes-bosns}
\begin{array}{rl}
D^{\mathrm{gh}\,\Phi\,\mathrm{NS}}_m=&i\left(\sum\limits_{r=1/2}^{m-1/2}\g_r\bt_{m-r}+\sum\limits_{r=1/2}^\infty\g_{m+r}\bt_{-r}+\sum\limits_{r=1/2}^\infty\g_{-r}\bt_{m+r}\right),\\[0.2cm]
D^{\mathrm{gh}\,\Phi\,\mathrm{NS}}_{-m}=&i\left(\sum\limits_{r=1/2}^{m-1/2}\g_{-r}\bt_{-(m-r)}+\sum\limits_{r=1/2}^\infty\g_{-(m+r)}\bt_{r}+\sum\limits_{r=1/2}^\infty\g_r\bt_{-(m+r)}\right).
\end{array}
\eeq

Taking into account above mode expansions of the extended constraints it is possible to check the nilpotency of the quantum BRST charge. In the case of $xp$-ordering for the space-time coordinates and momenta, Weyl-ordering for odd coordinates, $cb$- and $\g\bt$-ordering for ghosts BRST charge indeed appears to be nilpotent
\beq
\Omega^2|_{xp,\,\mathrm{Weyl}\!,\, cb-\mathrm{ordering}}=0
\eeq
extending the result of the previous section on the anomaly absence in the matter sector of the quantum constraint algebra. The choice of the ordering and nilpotency condition of the BRST charge fix the form of the zero-mode parts of the extended Virasoro
\beq
L^{\mathrm{ext}}_0=L^{\mathrm{m}}_0+\frac12L^{\mathrm{gh\, L}}_0+L^{\mathrm{gh\, T}}_0+L^{\mathrm{gh\, D}}_0+L^{\mathrm{gh}\,\Phi}_0
\eeq
and dilation generators
\beq
D^{\mathrm{ext}}_0=D^{\mathrm m}_0+D^{\mathrm{gh\, T}}_0+D^{\mathrm{gh}\,\Phi}_0.
\eeq
For the $xp$-ordering (Weyl-ordering for fermions) zero modes of the Virasoro generator for the matter variables were given in (\ref{vir0-mat-bos-yp}), (\ref{vir0-fermr-weyl}) and (\ref{vir0-fermns-weyl}). For contributions of the fermionic ghosts to the extended Virasoro generator we get
\beq
L^{\mathrm{gh\, T(D)}}_{0\, cb}=\sum\limits_{n=1}^\infty n(c^{T(D)}_{-n}b^{T(D)}_{n}-c^{T(D)}_{n}b^{T(D)}_{-n})
\eeq
and the same expression holds for the contribution of $\frac12L^{\mathrm{gh\, L}}_0$. For the bosonic ghosts depending on their periodicity one finds
\beq
L^{\mathrm{gh}\,\Phi\,\mathrm{R}}_{0\,\g\bt}=\sum\limits_{n=1}^\infty n(\g_{n}\bt_{-n}-\g_{-n}\bt_{n})\quad\mbox{or}\quad L^{\mathrm{gh}\,\Phi\,\mathrm{NS}}_{0\,\g\bt}=\sum\limits_{r=1/2}^\infty r(\g_{r}\bt_{-r}-\g_{-r}\bt_{r}).
\eeq
Similarly the contribution of the $xp$-ordered matter variables to the zero mode of the dilatation generator were obtained in (\ref{dil0-mat-yp}). Contribution of the $(c^T,b^T)$ ghosts is
\beq
D^{\mathrm{gh\, T}}_{0\, cb}=-2i\left(c^T_0b^T_0+\sum\limits_{n=1}^\infty(c^T_{n}b^T_{-n}+c^T_{-n}b^T_{n})\right)
\eeq
and that of $(\g,\bt)$ ghosts is
\beq
D^{\mathrm{gh}\,\Phi\, \mathrm{R}}_{0\,\g\bt}=i\left(\g_0\bt_0+\sum\limits_{n=1}^\infty(\g_{n}\bt_{-n}+\g_{-n}\bt_{n})\right)\quad\mbox{or}\quad D^{\mathrm{gh}\,\Phi\, \mathrm{NS}}_{0\,\g\bt}=i\sum\limits_{r=1/2}^\infty(\g_{r}\bt_{-r}+\g_{-r}\bt_{r}).
\eeq

In case of the $(-+)-$ordering quantum BRST charge is not nilpotent
\beq
\begin{array}{rl}
\Omega^2|_{(-+)-\mathrm{ordering}}=&\left(\frac{5(d+1)}{24}-\frac{43}{12}\right)\sum\limits_{n=1}^\infty n^3c^L_{-n}c^L_n+(d-2)\sum\limits_{n=1}^\infty n c^D_{-n}c^D_n\\[0.2cm]
+&\frac{i}{2}(d-3)\sum\limits_{n=1}^\infty n^2(c^D_{-n}c^L_n-c^L_{-n}c^D_n).
\end{array}
\eeq
The obstacle on the r.h.s. is given by three anomalous
contributions. The first infinite sum is proportional to the
Virasoro anomaly to which each $(X,P)$-pair contributes $+2$ and
each odd $\Xi$-variable contributes $+1/2$, $(c^L,b^L)$ and
$(c^T,b^T)$ ghosts each contribute --26, $(c^D,b^D)$ ghosts --2 and
$(\g,\bt)$ ghosts +11 in accordance with the standard Virasoro
anomaly calculus (see, e.g., \cite{Pol-book}). The second sum is
proportional to the dilatation anomaly. $(X,P)-$variables contribute to its value $+(d+1)$, $(\g,\bt)$ ghosts +1 and $(c^T,b^T)$
ghosts --4. The last term is proportional to the mixed
dilatation-Virasoro anomaly. If contribution of the
$(X,P)$-variables is normalized to $+(d+1)$, then that of $\g\bt$
ghosts is +2 and that of $(c^T,b^T)$ ghosts is --6. It is clear
that for such ordering not all anomalies can be canceled
suggesting a modification of the model by adding extra variables
and/or constraints.\footnote{In the recently proposed chiral
world-sheet CFT models \cite{AMP} (that we believe are also null
by their nature) with Minkowski target-space realized as a
projective light-cone embedded into flat space-time with extra
time-like and space-like dimensions the set of the constraints was
adjusted in such a way that dilatation and mixed anomalies have
the same value and can be simultaneously canceled fixing the value of the
space-time dimension.}

Let us finally remark that the above
expression for the square of the BRST charge corresponds to the
following form of the zero mode of the extended Virasoro generator
\beq
L^{\mathrm{ext}}_{0\,(-+)}=L^{\mathrm{m}}_{0\,(-+)}+\frac12L^{\mathrm{gh\,
L}}_{0\,(-+)}+L^{\mathrm{gh\, T}}_{0\,(-+)}+L^{\mathrm{gh\,
D}}_{0\,(-+)}+L^{\mathrm{gh}\,\Phi}_{0\,(-+)},
\eeq
where the
matter contributions have been defined in
(\ref{vir-bos-mat-zero-mode-frord}),
(\ref{vir-mat-fermr-zero-mode-frord}) and
(\ref{vir-mat-fermns-zero-mode-frord}). Contributions of the fermionic
ghosts read
\beq L^{\mathrm{gh\,
T(D)}}_{0\,(-+)}=\sum\limits_{n=1}^\infty
n(c^{T(D)}_{-n}b^{T(D)}_{n}+b^{T(D)}_{-n}c^{T(D)}_{n})+\frac{1}{12},
\eeq
with the same expression for $\frac12L^{\mathrm{gh\,
L}}_{0\,(-+)}$ and that of bosonic ghosts is
\beq
L^{\mathrm{gh}\,\Phi\,
\mathrm{R}}_{0\,(-+)}=\sum\limits_{n=1}^\infty
n(\bt_{-n}\g_n-\g_{-n}\bt_{n})-\frac{1}{12}\quad\mbox{or}\quad
L^{\mathrm{gh}\,\Phi\,
\mathrm{NS}}_{0\,(-+)}=\sum\limits_{r=1/2}^\infty
r(\bt_{-r}\g_r-\g_{-r}\bt_r)+\frac{1}{24}.
\eeq
The values of
numeric constants equal those prescribed for the (anti)periodic
bosons/fermions by the $\zeta-$function regularization of the relevant
infinite sums \cite{Pol-book}. Zero mode of the extended
dilatation generator for the $(-+)-$ordering
\beq
D^{\mathrm{ext}}_{0(-+)}=D^{\mathrm m}_{0(-+)}+D^{\mathrm{gh\,
T}}_{0(-+)}+D^{\mathrm{gh}\,\Phi}_{0(-+)}
\eeq
is defined by the matter term (\ref{dil-mat-zero-mode-frord}) and ghosts
\beq
D^{\mathrm{gh\,
T}}_{0(-+)}=-2i\left(c^T_0b^T_0+\sum\limits_{n=1}^\infty(c^T_{-n}b^T_n-b^T_{-n}c^T_n)-\frac12\right),
\eeq
\beq
D^{\mathrm{gh}\,\Phi\,\mathrm{R}}_{0(-+)}=i\left(\g_0\bt_0+\sum\limits_{n=1}^\infty(\g_{-n}\bt_n+\bt_{-n}\g_n)-\frac12\right)\quad\mbox{or}\quad
D^{\mathrm{gh}\,\Phi\,\mathrm{NS}}_{0(-+)}=i\sum\limits_{r=1/2}^\infty(\g_{-r}\bt_r+\bt_{-r}\g_r).
\eeq

\section{Conclusion and outlook}

Viewing anti-de Sitter space-time as a projective space combines
linearization of the $SO(2,d-1)$ isometry group action with the
absence of any restrictions on the homogeneous coordinates
parametrizing it which suggests an interesting perspective for studying
models of point-like and extended objects. Here our attention has
been restricted to the consideration of the simplest models of the
massless point particle and the null string with minimal world-line
(world-sheet) supersymmetry. At the classical level starting with
the phase-space representation of the action as the integral of the pullback
of the symplectic 1-form supplemented by the sum of the
first-class constraints with the Lagrange multipliers various forms of the massless point particle and null string
Lagrangians have been obtained
by integrating out the momenta and the part
of Lagrange multipliers and compared with
other models in which conformal symmetry is linearly realized.

Quantization of the spinning particle model has been shown to produce
the Dirac-type equation in homogeneous embedding-space
coordinates of the form similar to that of the Fang-Fronsdal equations for
the fermionic fields on $AdS_d$.

The form of the possible constraints for the null string is severely restricted by the compatibility with the world-sheet conformal
reparametrizations. So we have considered the realization of the $2d$ minimal supersymmetry algebra
extended by the space-time dilatation and Virasoro generators by the
quadratic constraints. At the quantum
level we examined two realizations of this classical algebra: one corresponding
to the generalization of the coordinate-momentum ordering and another
based on the ordering in terms of the positive/negative Fourier modes of the constituent
operators. These orderings are associated with two vacua
commonly used when quantizing null strings: one annihilated by
the momentum conjugate to the space-time coordinates and the other
annihilated by the positive modes of both coordinate and
momentum operators. In the case of the generalized coordinate-momentum
ordering BRST charge remains nilpotent, while for the ordering
in terms of the positive/negative Fourier modes the square of the
BRST charge is proportional to the sum of the Virasoro,
dilatation and mixed anomalies. Their cancelation necessitates a
modification of the model adding extra variables and/or constraints.

The fact that the quantum BRST charge is nilpotent in any space-time dimension for the generalized coordinate-momentum ordering invites further exploration of this case. Experience in quantizing tensionless string models on the flat and $AdS$ backgrounds \cite{Lizzi}, \cite{Zabzine}, \cite{Bonelli'03N}, \cite{Bonelli'03}, \cite{Sagnotti}, \cite{Bonelli'04} hints at the presence in the spectrum of the supermultiplets of massless higher-spin fields on  $AdS_d$ in the projective-space formulation.
For the ordering in terms of positive/negative Fourier modes of the operators  further examination of the associated theory (upon certain amendments to cancel Virasoro and dilatation anomalies) could consist of considering correlation functions of the vertices. In the ambitwistor string model \cite{Mason'13} and its predecessors \cite{Witten03}, \cite{Berkovits} which have a similar structure such correlators are known to reproduce tree-level scattering amplitudes for a variety of field theories on the flat background \cite{Ohmori}, \cite{CGMMR}. In our case it is tempting to suggest that they could give correlators for field theories on $AdS_d$ that in the boundary limit coincide with the dual free $CFT$\footnote{That the dual $CFT$ is expected to be a free field theory follows from the fact the tension is zero in our string model.} correlators, whose stringy interpretation has been recently probed in \cite{AMP}.

Other direction of extension of our results is to consider particle and string models for which the projective description of anti-de Sitter space-time combines with the space-time supersymmetry. It is known that in the case of $4-$dimensional anti-de Sitter space its $SO(2,3)$ isometry is the subgroup of $OSp(4|N)$ supergroup and $SO(2,4)\sim SU(2,2)$ isometry group of the $5-$dimensional anti-de Sitter space is the subgroup of $SU(2,2|N)$. For flat 4-dimensional space-time it is well known that linear realization of $SU(2,2)$ and $SU(2,2|N)$ is achieved in the framework of the (super)twistor theory \cite{Penrose}, \cite{Ferber}\footnote{For the embedding space description of $D=4$ superconformal theories that uses supertwistors see, e.g., \cite{Siegel}, \cite{Skiba}.}  and for the higher-dimensional spaces linear realization of respective (generalized) superconformal symmetries using higher-dimensional generalizations of supertwistors \cite{BL'98}, \cite{BLS}, \cite{BdAPV}, \cite{Bars2}, \cite{BdAS}, \cite{U'07}. Generalization of $SU(2,2)$ twistors (actually ambitwistors) to the case of $AdS_5$ space was recently considered in \cite{Williams} and it is quite plausible that the above mentioned higher-dimensional (super)twistors can be applied to the linearization of isometries of the higher-dimensional $AdS$ spaces compatible the space-time supersymmetry.\footnote{For various previous approaches to the construction of (super)twistors for $AdS$ spaces see, e.g., \cite{quarks}, \cite{Claus}, \cite{BLPS}, \cite{Cederwall} and also \cite{U'16}, \cite{Townsend}.}


\begin{thebibliography}{99}
\bibitem{Dirac}
P.A.M.~Dirac, The electron wave equation in de-Sitter space, Ann. Math. \textbf{36} (1935) 657; Wave equations in conformal space, Ann. Math. \textbf{37} (1936) 429.
\bibitem{MN-1}
R.~Marnelius and B.E.W.~Nilsson, Equivalence between a massive spinning particle in Minkowski space and a massless one in a de~Sitter space, Phys. Rev. \textbf{D20} (1979) 839.
\bibitem{Marnelius}
R.~Marnelius, Manifestly conformally  covariant description of spinning and charged particles, Phys. Rev. \textbf{D20} (1979) 2091.
\bibitem{Bars}
I.~Bars, C.~Deliduman and O.~Andreev, Gauged duality, conformal symmetry and space-time with two times, Phys. Rev. \textbf{D58} (1998) 066004; \href{http://arxiv.org/abs/hep-th/9803188}{arXiv:hep-th/9803188};\\
I.~Bars, C.~Deliduman and D.~Minic, Strings, branes and two-time physics, Phys. Lett. \textbf{B466} (1999) 135; \href{http://arxiv.org/abs/hep-th/9906223}{arXiv:hep-th/9906223}.
\bibitem{Salam}
G.~Mack and A.~Salam, Finite component field representations of the conformal group, Ann. Phys. \textbf{53} (1969) 174.
\bibitem{Ferrara}
S.~Ferrara, R.~Gatto and A.F.~Grillo, Conformal algebra in space-time and operator product expansion, Springer Tracts Mod. Phys. \textbf{67} (1973) 1.
\bibitem{Martensson}
U.~M\r artensson, The spinning conformal particle and its BRST quantization, Int. J. Mod. Phys. \textbf{A8} (1993) 5305.
\bibitem{GLSSV-conf-string}
H.~Gustafsson, U.~Lindstr\" om, P.~Saltsidis, B.~Sundborg and R.~von Unge, Hamiltonian BRST quantization of the conformal string, Nucl. Phys. \textbf{B440} (1995) 495; \href{http://arxiv.org/abs/hep-th/9410143}{arXiv:hep-th/9410143}.
\bibitem{Saltsidis-conf-string}
P.~Saltsidis, Hamiltonian BRST quantization of the conformal spinning string, Nucl. Phys. \textbf{B446} (1995) 286; \href{http://arxiv.org/abs/hep-th/9503062}{arXiv:hep-th/9503062}.

\bibitem{CCP}
L.~Cornalba, M.S.~Costa and J.~Penedones, Deep inelastic scattering in conformal QCD, JHEP \textbf{1003} (2010) 133; \href{http://arxiv.org/abs/0911.0043}{arXiv:0911.0043 [hep-th]}.
\bibitem{Weinberg}
S.~Weinberg, Six-dimensional methods for four-dimensional conformal field theories, Phys. Rev. \textbf{D82} (2010) 045031; \href{http://arxiv.org/abs/1006.3480}{arXiv:1006.3480 [hep-th]}.
\bibitem{Penedones}
J.~Penedones, Writing CFT correlation functions as AdS scattering amplitudes, JHEP \textbf{1103} (2011) 025; \href{http://arxiv.org/abs/1011.1485}{arXiv:1011.1485 [hep-th]}.

\bibitem{Grig}
X.~Bekaert and M.~Grigoriev, Notes on the ambient approach to boundary values of $AdS$ gauge fields, J. Phys. \textbf{A46} (2013) 214008; \href{http://arxiv.org/abs/1207.3439}{arXiv:1207.3439  [hep-th]}.
\bibitem{Didenko}
V.E.~Didenko and E.D.~Skvortsov, Towards higher-spin holography in ambient space of any dimension, J. Phys. \textbf{A46} (2013) 214010; \href{http://arxiv.org/abs/1207.6786}{arXiv:1207.6786 [hep-th]}.
\bibitem{Williams}
T.~Adamo, D.~Skinner and J.~Williams, Twistor methods for $AdS_5$, JHEP \textbf{1608} (2016) 167; \href{http://arxiv.org/abs/1607.03763}{arXiv:1607.03763 [hep-th]}.

\bibitem{Bonelli'03}
G.~Bonelli, On the covariant quantization of tensionless bosonic strings in $AdS$ spacetime, JHEP {\bf 0311} (2003) 028; \href{http://arxiv.org/abs/hep-th/0309222}{arXiv:hep-th/0309222}.

\bibitem{Rivelles}
L.A.~Cabral and V.O.~Rivelles, Particles and strings in degenerate metric spaces, Class. Quantum Grav. \textbf{17} (2000) 1577; \href{http://arxiv.org/abs/hep-th/99101633}{arXiv:hep-th/9910163}.

\bibitem{Fronsdal}
C.~Fronsdal, Singletons and massless, integral-spin fields on de Sitter space, Phys. Rev. \textbf{D20} (1979) 848.
\bibitem{FangF}
J.~Fang and C.~Fronsdal, Massless, half-integer-spin fields in de Sitter space, Phys. Rev. \textbf{D22} (1980) 1361.
\bibitem{Metsaev'95}
R.~Metsaev, Massless mixed symmetry bosonic free fields in $d$-dimensional anti-de Sitter space-time, Phys. Lett. \textbf{B354} (1995) 78.
\bibitem{Metsaev'97}
R.R.~Metsaev, Fermionic fields in the $d-$dimensional anti-de Sitter spacetime, Phys. Lett. \textbf{B419} (1998) 49; \href{http://arxiv.org/abs/hep-th/9802097}{arXiv:hep-th/9802097}.

\bibitem{Brink}
L.~Brink and M.~Henneaux, Principles of string theory, Plenum Press, N.Y., 1988, 297p.

\bibitem{Mcfarlane}
A.C.~Davis, A.J.~Macfarlane, P.C.~Popat and J.W.~van Holten, The quantum mechanics of the supersymmetric nonlinear $\s-$model, J. Phys. \textbf{A17} (1984) 2945.\\
A.J.~Macfarlane and P.C.~Popat, The quantum mechanics of the $N=2$ extended supersymmetric nonlinear $\s-$model, J. Phys. \textbf{A17} (1984) 2955.
\bibitem{Smilga}
A.V.~Smilga, How to quantize supersymmetric theories, Nucl. Phys. \textbf{B292} (1987) 363.

\bibitem{LST-null-spin-string}
U.~Lindstr\" om, B.~Sundborg and G.~Theodoridis, The zero tension limit of the spinning string, Phys. Lett. \textbf{B258} (1991) 331.
\bibitem{Karlhede}
A.~Karlhede and U.~Lindstr\" om, The classical bosonic string in the zero tension limit, Class. Quantum Grav. \textbf{3} (1986) L73.
\bibitem{AAZ'87}
A.A.~Zheltukhin, Hamiltonian structure of the antisymmetric action of a string, JETP Lett. \textbf{46} (1987) 262 [Pisma v ZhETF \textbf{46} (1987) 208].

\bibitem{Pol-book}
J.~Polchinski, String theory. V.1. An introduction to the bosonic string. Cambridge University Press, 1998, 402p.

\bibitem{BZh}
I.A.~Bandos and A.A.~Zheltukhin, Hamiltonian mechanics and absence of critical dimensions for null membranes, Sov. J. Nucl. Phys.  {\bf 50} (1989) 556 [Yad. Fiz.  {\bf 50} (1989) 893];\\
Quantum theory of closed null supermembranes in four-dimensional space, JETP Lett.  {\bf 53} (1991) 5 [Pisma v ZhETF  {\bf 53} (1991) 7];\\
Null super p-brane: Hamiltonian dynamics and quantization, Phys. Lett. {\bf B261} (1991) 245;\\
Covariant quantization of null supermembranes in four-dimensional space-time, Theor. Math. Phys.  {\bf 88} (1991) 925 [Teor. Mat. Fiz.  {\bf 88} (1991) 358].
\bibitem{ILST'93}
J.~Isberg, U.~Lindstr\" om, B.~Sundborg and G.~Theodoridis, Classical and quantized tensionless strings, Nucl. Phys. {\bf B411} (1994) 122; \href{http://arxiv.org/abs/hep-th/9307108}{arXiv:hep-th/9307108}.
\bibitem{GRRA}
J.~Gamboa, C.~Ramirez and M.~Ruiz-Altaba, Null Spinning Strings, Nucl. Phys. {\bf B338} (1990) 143.

\bibitem{Mason'13}
L.~Mason and D.~Skinner, Ambitwistor strings and the scattering equations, JHEP \textbf{1407} (2014) 048; \href{http://arxiv.org/abs/1311.2564}{arXiv:1311.2564 [hep-th]}.

\bibitem{Tourkine}
E.~Casali and P.~Tourkine, On the null origin of the ambitwistor string, JHEP {\bf 1611} (2016) 036; \href{http://arxiv.org/abs/1606.05636}{arXiv:1606.05636 [hep-th]}.
\bibitem{Bandos'14}
I.~Bandos, Twistor/ambitwistor strings and null-superstrings in space-time of $D=4,10$ and $11$ dimensions, JHEP \textbf{1409} (2014) 086; \href{http://arxiv.org/abs/1404.1299}{arXiv:1404.1299 [hep-th]}.
\bibitem{BdAM}
I.A.~Bandos, J.A.~de Azcarraga and C.~Miquel-Espanya, Superspace formulations of the (super)twistor string, JHEP \textbf{0607} (2006) 005; \href{http://arxiv.org/abs/hep-th/0604037}{arXiv:hep-th/0604037}.


\bibitem{Lizzi}
F.~Lizzi, B.~Rai, G.~Sparano and A.~Srivastava, Quantization of the null string and absence of critical dimensions, Phys. Lett. \textbf{B182} (1986) 326.
\bibitem{Zabzine}
U.~Lindstr\" om and M.~Zabzine, Tensionless strings, WZW models at critical level and massless higher spin fields, Phys. Lett. \textbf{B584} (2004) 178; \href{http://arxiv.org/abs/hep-th/0305098}{arXiv:hep-th/0305098}.
\bibitem{Bonelli'03N}
G.~Bonelli, On the tensionless limit of bosonic strings, infinite symmetries and
higher spins, Nucl.Phys. \textbf{B669} (2003) 159; \href{http://arxiv.org/abs/hep-th/0305155}{arXiv:hep-th/0305155}.
\bibitem{Sagnotti}
A.~Sagnotti and M.~Tsulaia, On higher spins and the tensionless limit of string theory, Nucl. Phys. \textbf{B682} (2004) 83; \href{http://arxiv.org/abs/hep-th/0311257}{arXiv:hep-th/0311257}.
\bibitem{Bonelli'04}
G.~Bonelli, On the boundary gauge dual of closed tensionless free strings in AdS, JHEP \textbf{0411} (2004) 059; \href{http://arxiv.org/abs/hep-th/0407144}{arXiv:hep-th/0407144}.

\bibitem{Witten03}
E.~Witten, Perturbative gauge theory as a string theory in twistor space, Commun. Math. Phys. \textbf{252} (2004) 189; \href{http://arxiv.org/abs/hep-th/0312171}{arXiv:hep-th/0312171}.
\bibitem{Berkovits}
N.~Berkovits, Alternative string theory in twistor space for $N=4$ super-Yang-Mills theory, Phys. Rev. Lett. \textbf{93} (2004) 011601; \href{http://arxiv.org/abs/hep-th/0402045}{arXiv:hep-th/0402045}.
\bibitem{Ohmori}
K.~Ohmori, Worldsheet geometries of ambitwistor string, JHEP \textbf{1506} (2015) 075; \href{http://arxiv.org/abs/1504.02675}{arXiv:1504.02675 [hep-th]}.
\bibitem{CGMMR}
E.~Casali, Y.~Geyer, L.~Mason, R.~Monteiro and K.A.~Roehrig, New ambitwistor string theories, JHEP \textbf{1511} (2015) 038; \href{http://arxiv.org/abs/1506.08771}{arXiv:1506.08771 [hep-th]}.


\bibitem{AMP}
T.~Adamo, R.~Monteiro and M.F.~Paulos, Space-time CFTs from the Riemann sphere, JHEP \textbf{1708} (2017) 067; \href{http://arxiv.org/abs/1703.04589}{arXiv:1703.04589 [hep-th]}.

\bibitem{Penrose}
R.~Penrose,  Twistor algebra,  J.\ Math.\ Phys.\  {\bf 8} (1967) 345.
\bibitem{Ferber}
A.~Ferber, Supertwistors and Conformal Supersymmetry,  Nucl.\ Phys.\ {\bf B132} (1978) 55.

\bibitem{Siegel}
W.~Siegel, AdS/CFT in superspace, \href{http://arxiv.org/abs/1005.2317}{arXiv:1005.2317 [hep-th]}.\\
Embedding versus $6D$ twistors, \href{http://arxiv.org/abs/1204.5679}{arXiv:1204.5679 [hep-th]}.
\bibitem{Skiba}
W.D.~Goldberger, W.~Skiba and M.~Son, Superembedding methods for
$4d$ $\mc N=1$ SCFTs, Phys. Rev. \textbf{D86} (2012) 025019,
\href{http://arxiv.org/abs/1112.0325}{arXiv:1112.0325 [hep-th]}.

\bibitem{BL'98}
I.~Bandos and J.~Lukierski, Tensorial central charges and
new superparticle models with fundamental spinor coordinates, Mod.
Phys. Lett. {\bf A14} (1999) 1257; \href{http://arxiv.org/abs/hep-th/9811022}{arXiv:hep-th/9811022}.
\bibitem{BLS}
I.A. Bandos, J. Lukierski and D.P. Sorokin, Superparticle
models with tensorial central charges, Phys. Rev. {\bf D61}(2000)
045002; \href{http://arxiv.org/abs/hep-th/9904109}{arXiv:hep-th/9904109}.
\bibitem{BdAPV}
I.A.~Bandos, J.A.~de Azcarraga, M.~Picon and O.~Varela, Supersymmetric string model with 30 $\kappa-$symmetries in an extended $D=11$ superspace and $\frac{30}{32}$ BPS states, Phys. Rev. {\bf D69} (2004) 085007; \href{http://arxiv.org/abs/hep-th/0307106}{arXiv:hep-th/0307106}.
\bibitem{Bars2}
I.~Bars and M.~Picon, Twistor transform in $d$
dimensions and a unifying role for twistors, Phys. Rev. {\bf D73}
(2006) 064033; \href{http://arxiv.org/abs/hep-th/0512348}{arXiv:hep-th/0512348}.
\bibitem{BdAS}
I.A.~Bandos, J.A.~de Azcarraga and D.~Sorokin, On $D=11$
supertwistors, superparticle quantization and a hidden $SO(16)$
symmetry of supergravity; \href{http://arxiv.org/abs/hep-th/0612252}{arXiv:hep-th/0612252}.
\bibitem{U'07}
D.V.~Uvarov, Supertwistor formulation for higher dimensional superstrings, Class. Quantum Grav. {\bf24} (2007) 5383; \href{http://arxiv.org/abs/hep-th/0703051}{arXiv:hep-th/0703051}.
\bibitem{quarks}
P.~Claus, M.~Gunaydin, R.~Kallosh, J.~Rahmfeld and Y.~Zunger, Supertwistors as quarks of $SU(2, 2|4)$, JHEP {\bf 9905} (1999) 019; \href{http://arxiv.org/abs/hep-th/9905112}{arXiv:hep-th/9905112}.
\bibitem{Claus}
P.~Claus, R.~Kallosh and J.~Rahmfeld, BRST quantization of a particle in $AdS_5$,
 Phys.\ Lett.\ {\bf B462} (1999) 285; \href{http://arxiv.org/abs/hep-th/9906195}{arXiv:hep-th/9906195}.
\bibitem{BLPS}
I.A.~Bandos, J.~Lukierski, C.~Preitschopf and D.P.~Sorokin, $OSp$ supergroup manifolds, superparticles and supertwistors,  Phys.\ Rev.\ {\bf D61} (2000) 065009; \href{http://arxiv.org/abs/hep-th/9907113}{arXiv:hep-th/9907113}.
\bibitem{Cederwall}
M.~Cederwall, Geometric construction of $AdS$ twistors, Phys.\ Lett.\ {\bf B483} (2000) 257; \href{http://arxiv.org/abs/hep-th/0002216}{arXiv:hep-th/0002216}.\\
AdS twistors for higher spin theory,  AIP Conf.\ Proc.\  {\bf 767} (2005) 96; \href{http://arxiv.org/abs/hep-th/0412222}{arXiv:hep-th/0412222}.
\bibitem{U'16}
D.V.~Uvarov, Ambitwistors, oscillators and massless fields on $AdS_5$, Phys. Lett. \textbf{B762} (2016) 415; \href{http://arxiv.org/abs/1607.05233}{arXiv:1607.05233 [hep-th]}.
\bibitem{Townsend}
A.S.~Arvanitakis, A.E.~Barns-Graham and P.K.~Townsend, Anti-de Sitter particles and manifest (super)isometries, Phys. Rev. Lett. \textbf{118} (2017) 141601; \href{http://arxiv.org/abs/1608.04380}{arXiv:1608.04380 [hep-th]}.
\end{thebibliography}
\end{document}